\documentclass[sigconf]{acmart}
\acmSubmissionID{946}

\usepackage{xcolor}
\usepackage{enumitem}
\usepackage{subfig}
\usepackage{subcaption}
\usepackage{xspace}
\copyrightyear{2026}
\acmYear{2026}
\setcopyright{cc}
\setcctype{by}
\acmConference[SoCC '26 Vol. 2]{ACM Symposium on Cloud Computing Volume 2}{November 18--20, 2026}{Singapore, Singapore}
\acmBooktitle{ACM Symposium on Cloud Computing Volume 2 (SoCC '26 Vol. 2), November 18--20, 2026, Singapore, Singapore}
\acmDOI{10.1145/3828161.3857887}
\acmISBN{979-8-4007-2816-7/2026/11}
\begin{document}

\title{Rethinking AI Cloud Infrastructure for Agentic Serving Systems with the Aries\xspace Experimentation Framework}

\author{JooYoung Park}
\authornote{These authors contributed equally as co-first authors.}
\orcid{0000-0002-6979-9362}
\affiliation{\institution{NTU Singapore}
\city{Singapore}
\country{Singapore}}
\email{jooyoung001@e.ntu.edu.sg}

\author{Leonid Kondrashov}
\authornotemark[1]\orcid{0009-0003-1090-9437}
\affiliation{\institution{NTU Singapore}
\city{Singapore}
\country{Singapore}}
\email{leonid001@e.ntu.edu.sg}

\author{Boxi Zhou}
\authornotemark[1]\orcid{0009-0002-4619-2151}
\affiliation{\institution{NTU Singapore}
\city{Singapore}
\country{Singapore}}
\email{bzhou011@e.ntu.edu.sg}

\author{Hongrui Liu}
\authornotemark[1]\orcid{0009-0008-8026-3930}
\affiliation{\institution{NTU Singapore}
\city{Singapore}
\country{Singapore}}
\email{hongrui001@e.ntu.edu.sg}

\author{Zonghao Liu}
\orcid{0009-0004-4168-5533}
\affiliation{\institution{NTU Singapore}
\city{Singapore}
\country{Singapore}}
\email{liuz0138@e.ntu.edu.sg}

\author{Chengzhi Lu}
\authornote{Corresponding author.}\orcid{0000-0002-6421-3828}
\affiliation{\institution{NTU Singapore}
\city{Singapore}
\country{Singapore}}
\email{chengzhi.lu@ntu.edu.sg}

\author{Riccardo Mancini}
\orcid{0000-0002-1053-9400}
\affiliation{\institution{AWS}
\city{Cambridge}
\country{United Kingdom}}
\email{mancio@amazon.co.uk}

\author{Esha Choukse}
\orcid{0000-0003-0371-5522}
\affiliation{\institution{Microsoft}
\city{Seattle}
\state{Washington}
\country{USA}}
\email{esha.choukse@microsoft.com}

\author{Haris Javaid}
\orcid{0009-0008-3472-0803}
\affiliation{\institution{AMD}
\city{Singapore}
\country{Singapore}}
\email{haris.javaid@amd.com}

\author{German Sviridov}
\orcid{0000-0002-7814-3330}
\affiliation{\institution{AMD}
\city{Singapore}
\country{Singapore}}
\email{German.Sviridov@amd.com}

\author{Tao Peng}
\orcid{0009-0002-4805-0237}
\affiliation{\institution{Ant Group}
\city{Hangzhou}
\country{China}}
\email{bergwolf@antgroup.com}

\author{Chen Zhao}
\orcid{0009-0006-8871-0323}
\affiliation{\institution{Ant Group}
\city{Hangzhou}
\country{China}}
\email{winters.zc@antgroup.com}

\author{Marios Kogias}
\orcid{0009-0006-7034-5284}
\affiliation{\institution{Imperial College London}
\city{London}
\country{United Kingdom}}
\email{m.kogias@imperial.ac.uk}

\author{Luo Mai}
\orcid{0000-0002-3594-1092}
\affiliation{\institution{University of Edinburgh}
\city{Edinburgh}
\country{United Kingdom}}
\email{luo.mai@ed.ac.uk}

\author{Anastasia Avdeeva}
\orcid{0000-0001-7368-6099}
\affiliation{\institution{NCSpeech}
\city{Newark}
\country{USA}}
\email{ananaskelly@ncspeech.org}

\author{Aleksei Gusev}
\orcid{0009-0001-5200-747X}
\affiliation{\institution{NCSpeech}
\city{Newark}
\country{USA}}
\email{aleksei.gusev@ncspeech.org}

\author{Dmitrii Ustiugov}
\orcid{0000-0003-3156-010X}
\affiliation{\institution{NTU Singapore}
\city{Singapore}
\country{Singapore}}
\email{dmitrii.ustiugov@ntu.edu.sg}

% \settopmatter{authorsperrow=1}
% \author{
% JooYoung Park\textsuperscript{*1},
% Leonid Kondrashov\textsuperscript{*1},
% Boxi Zhou\textsuperscript{*1},
% Hongrui Liu\textsuperscript{*1},
% Zonghao Liu\textsuperscript{1},
% Chengzhi Lu\textsuperscript{$\dagger$1},
% Riccardo Mancini\textsuperscript{2},
% Esha Choukse\textsuperscript{3},
% Haris Javaid\textsuperscript{4},
% German Sviridov\textsuperscript{4},
% Tao Peng\textsuperscript{5},
% Chen Zhao\textsuperscript{5},
% Anastasia Avdeeva\textsuperscript{6},
% Aleksei Gusev\textsuperscript{6},
% Marios Kogias\textsuperscript{7},
% Luo Mai\textsuperscript{8},
% Dmitrii Ustiugov\textsuperscript{1}
% }

% \affiliation{%
%  \institution{\textsuperscript{1}NTU Singapore,
%  \textsuperscript{2}AWS,
%  \textsuperscript{3}Microsoft,
%  \textsuperscript{4}AMD,
%  \textsuperscript{5}Ant Group,
%  \textsuperscript{6}NCSpeech,
%  \textsuperscript{7}Imperial College London,
%  \textsuperscript{8}University of Edinburgh}
%  \country{}
% }

\renewcommand{\shortauthors}{Park et al.}

\begin{abstract}
Autonomous agents challenge conventional LLM serving with long-lived, cross-component execution and evolving system state. These characteristics complicate controlled systems experimentation and attribution of end-to-end behavior across the serving stack.
We present Aries\xspace, a full-stack experimentation framework that separates task {specifications} from execution configurations, reconstructs cross-component agent trajectories with correlated system telemetry, and exposes stateful tool execution through a consistent interface across heterogeneous sandbox substrates. We use Aries\xspace to conduct reproducible experiments on open agent harnesses and benchmarks. We complement these experiments with production traces from a commercial platform, grounding low-level systems research in observed production behavior. Our results show that (1) token-centric metrics miss non-inference bottlenecks, (2) retaining additional context yields diminishing accuracy benefits while reducing serving capacity, and (3) tool sandboxes alternate between long idle periods and short resource bursts, while current snapshot-based state management makes aggressive suspension costly. 
% {A complementary threat-model discussion} further highlights the need to reduce the sandbox attack surface. 
We then discuss the vision for agent-native serving systems designed around trajectory-level metrics, adaptive context management, and elastic sandbox resource management.
% , and sandboxes with minimized attack surface.

\end{abstract}

\begin{CCSXML}
<ccs2012>
<concept>
<concept_id>10010520.10010521.10010537.10003100</concept_id>
<concept_desc>Computer systems organization~Cloud computing</concept_desc>
<concept_significance>500</concept_significance>
</concept>
<concept>
<concept_id>10010147.10010178.10010219.10010221</concept_id>
<concept_desc>Computing methodologies~Intelligent agents</concept_desc>
<concept_significance>500</concept_significance>
</concept>
</ccs2012>
\end{CCSXML}

\ccsdesc[500]{Computer systems organization~Cloud computing}
\ccsdesc[500]{Computing methodologies~Intelligent agents}

\keywords{AI agents, LLM serving, cloud infrastructure, tool sandboxes, benchmarking}

\maketitle

%\begingroup
%\renewcommand{\thefootnote}{}
%\footnotetext{*These authors contributed equally as co-first authors.  $\dagger$Corresponding author.}
%\endgroup

\vspace{-.7em}
\section{Introduction}

Large Language Models (LLMs) are rapidly evolving from prompt responders into autonomous agents. Open-source agents such as \textsc{OpenClaw} and \textsc{Hermes} have collectively attracted over half a million GitHub stars within months of their public releases~\cite{openclaw2026,hermesagent2026}, while closed-source commercial agents such as Claude and Codex are used by millions of people~\cite{codex}. An agent places the model inside a feedback loop: it observes the current state, selects an action, executes that action through a tool, and uses the result to decide how the task should continue~\cite{luo2025large,luo2026agentix,yao2022react}. This shifts the system problem from serving isolated model requests to sustaining end-to-end agent trajectories. Agent serving therefore no longer ends when the model emits tokens; each model response is only one step in an execution loop whose progress depends on both generated text and runtime state.

\textbf{Agent Serving Goes Beyond Inference.}
Like conventional LLM serving, agentic workloads are often hosted in cloud AI infrastructure~\cite{chen2025kairos,luo2026agentix}, but they impose fundamentally different system requirements. 
% however, featuring radically different system implications. 
Conventional LLM systems optimize individual model calls:
scheduling and resource management are focused on producing the next model response efficiently~\cite{kwon2023pagedattention,zheng2024sglang,yu2022orca,zhong2024distserve}. This abstraction works when the surrounding application logic remains outside the serving system. In agent serving, however, each model call is only one step in a task loop. Faster GPU decoding remains useful, but end-to-end progress also depends on retained context, tool execution, and sandbox management. Agent serving therefore requires a holistic, full-stack system design, beyond just fast inference serving.

\textbf{Task Progress Needs System Visibility.}
Existing evaluation methods capture only part of agent execution. Agent benchmarks measure whether a task reaches the correct outcome~\cite{padigela2025ml,zhou2024webarena,merrill2026terminal,liu2024agentbench}, whereas system benchmarks typically measure isolated model calls, stateless functions, or task transactions~\cite{reddi2020mlperf,ustiugov2021vhive,ferdman2012cloudsuite}. Neither of the above connects task progress and correctness to cross-layer runtime behavior, while {coupling task setup and evaluation to specific execution configurations} and substrate-dependent tool behavior hinder controlled comparison.
\setlength{\skip\footins}{1em}

We use the \textit{agent trajectory} as the unit of observation. A trajectory is the ordered record of model invocations, tool interactions, harness decisions, and the final task outcome. Based on this abstraction, we introduce \textit{Aries\xspace}, a {modular and configurable} experimentation framework that separates {benchmark-defined task setup and evaluation} from execution configuration, reconstructs cross-component trajectories with correlated telemetry, and standardizes tool execution across sandbox substrates, enabling reproducible comparisons and stage-level attribution of stalls and failures.
We use Aries\xspace to conduct reproducible experiments on open benchmarks and complement them with production traces from the AntGroup cluster, thereby grounding low-level systems research in observed production behavior.\footnote{The Aries\xspace code, toolchain, and production traces are available at \url{https://github.com/hyscale-lab/ARIES}.}

Our experiments reveal limitations that the conventional model-call metrics hide. Harness and tool execution can account for a heavy part of the end-to-end latency (e.g., up to 48\%), so LLM-engine token throughput may diverge from task-level execution efficiency.
Context retention creates an accuracy--capacity trade-off: controlled sweeps show that avoiding context overflows can raise task success from 55\% to 95\%, but gains plateau at workload-dependent thresholds, while production traces and controlled telemetry show that retained state reduces serving capacity. Production and controlled measurements also reveal mostly idle-but-bursty tool sandboxes, for which snapshot-based suspension remains economically infeasible. 
% {A complementary threat-model discussion} further motivates sandbox designs with minimal attack surfaces. 
Based on these insights, we discuss the vision for agent-native serving systems that feature trajectory-aware observability and control planes for trajectory context management and for the elasticity of agent-aware sandboxes.
% , and sandbox designs that minimize the attack surface of tool execution.

\vspace{-1em}
\section{Background and Motivation}
\vspace{-.2em}
\label{sec:background}
\begin{figure}
    \centering
    \vspace{-1em}
    \includegraphics[width=1\linewidth]{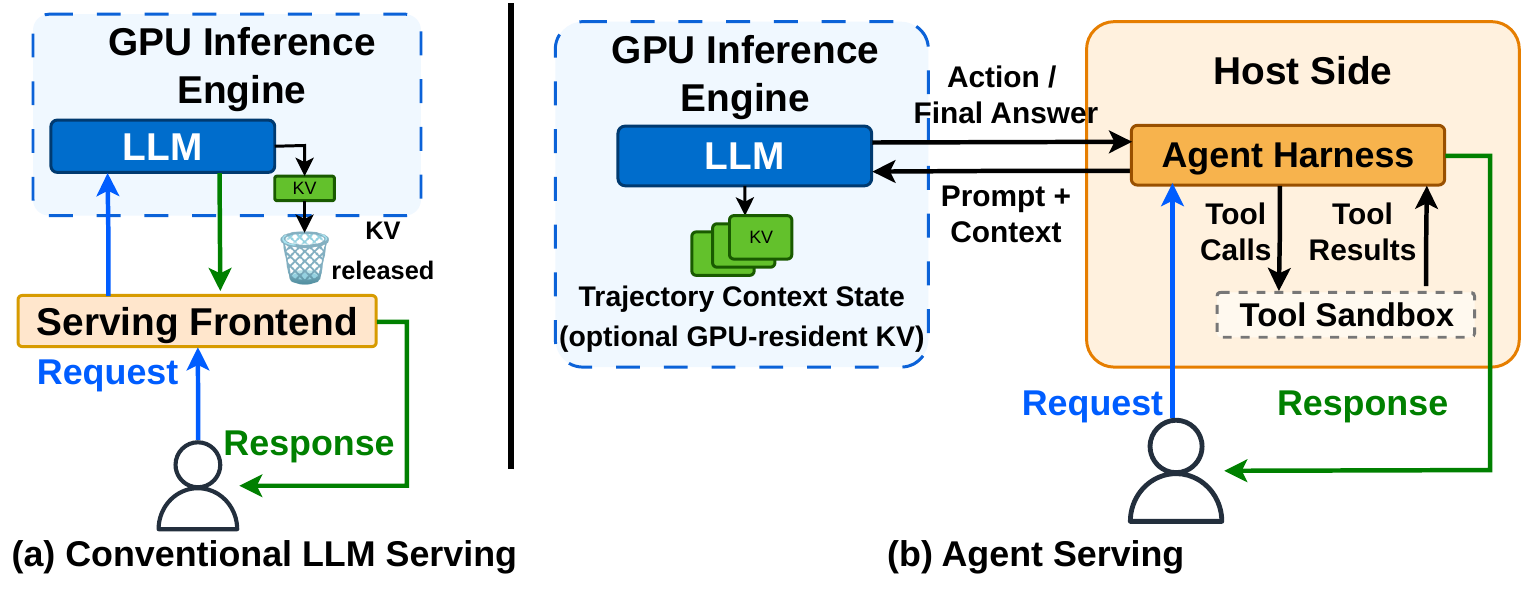}
    \vspace{-2.5em}
    \caption{Overview of conventional LLM-serving and emerging agent-serving systems.
    }
    \vspace{-1.5em}
    \label{fig:llm_vs_agent}
\end{figure}

\subsection{From Model Requests to Agent Trajectories}
\label{sec:agent-trajectory}
Conventional LLM serving systems are organized around the execution of individual model requests. As shown in Fig.~\ref{fig:llm_vs_agent}a, the serving runtime processes a prompt, generates output tokens, manages accelerator memory, and returns a response~\cite{zhong2024distserve,yu2022orca}. Scheduling and memory management are therefore scoped primarily to the active invocation. Depending on the cache policy, the engine may retain, offload, reuse, or reclaim the invocation’s KV-cache state after execution~\cite{kwon2023pagedattention,pan2025kvflow,zheng2024sglang,qin2024mooncake}. This request-centric design naturally prioritizes metrics such as time to first token, time per output token, request latency, and aggregate token throughput~\cite{kwon2023pagedattention,zhong2024distserve,patel2024splitwise,hu2024tetriinfer,yu2022orca,zheng2024sglang}.

Agent serving departs from this invocation-centric abstraction.
As shown in Fig.~\ref{fig:llm_vs_agent}b, the harness repeatedly invokes the LLM, dispatches tool calls to a sandboxed runtime, incorporates tool results into later prompts, and decides when the task is completed~\cite{luo2026agentix,liu2024agentbench,yao2022react}. A single user request may therefore span many dependent model and tool steps.

In this paper, we define an \textit{agent trajectory}
as the ordered sequence of LLM invocations, tool calls, tool results, and harness decisions produced while serving one user request.
Individual model requests are steps within this larger execution unit rather than independent transactions.
Trajectory state persists across these steps. The harness may represent this state as message history or structured events, while the serving layer may materialize reusable portions as cached model state, such as KV-cache prefixes. At the same time, model-generated actions execute on the host side within an isolated sandbox. Agent serving consequently spans GPU inference, host-side orchestration, persistent context state, and sandboxed execution, with dependencies across all four components.

\vspace{-1em}
\subsection{Limitations of Existing Benchmarking Frameworks}
\label{sec:benchmark-limitations}

Existing benchmarks answer complementary questions at incompatible granularities. Capability-oriented benchmarks evaluate whether an agent completes a realistic task, but largely treat the execution path as a black box~\cite{zhou2024webarena,gur2024real,jimenez2024swebench,merrill2026terminal,mialon2024gaia,xie2024osworld,yao2024tau}.
Systems benchmarks can measure end-to-end performance and resource behavior, but typically consider individual requests or transactions as their unit of analysis~\cite{reddi2020mlperf,ferdman2012cloudsuite,gan2019deathstarbench,ustiugov2021vhive}.

None of the prior frameworks and benchmarks reveal how system behavior affects progress within a multi-step agent task. Task setup and benchmark-specific evaluation are often coupled to harness- and runtime-specific execution configurations, hindering controlled comparisons. For example, OpenHands' SWE-bench integration uses task-specific images tied to its SDK revision and an evaluation wrapper that extracts patches from OpenHands-specific result records. Reusing the same benchmark across execution stacks therefore requires compatible task environments and evaluator-input adapters. Execution events remain fragmented across model, harness, and sandbox components, and stateful tool behavior depends on the underlying substrate. Agent serving therefore creates a measurement gap between {task progress} and cross-layer system execution, making it difficult to isolate configuration effects, reconstruct complete trajectories, and attribute resource bottlenecks to task-level delays and failures. 

\vspace{-.7em}
\section{Aries\xspace: Trajectory-Level Telemetry}
\label{sec:eval-setup}

\begin{figure}[t]
    \centering
    \vspace{-1em}
    \includegraphics[width=0.9\linewidth]{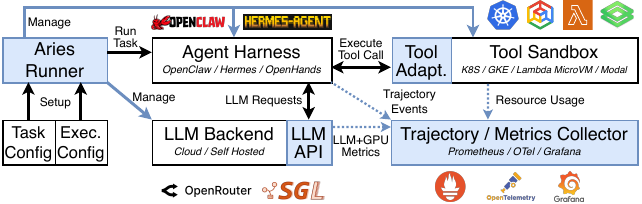}
    \vspace{-1.5em}
    \caption{Aries\xspace architecture.
    The blue-colored components are instrumentation in the agent-serving loop, dedicated to bridging the measurement gap.
    }
    \vspace{-2em}
    \label{fig:benchsuite_arch}
\end{figure}

The goal of Aries\xspace is to bridge the measurement gap between semantic task outcomes and cross-layer system behavior by treating an agent trajectory as the first-class citizen. This entails three challenges. First,
controlled experiments must preserve the task and its success criterion while varying execution configurations. Second, agent execution spans independently managed components, making trajectories difficult to reconstruct {and system effects difficult to attribute} from local logs. Third, tool execution is both stateful and substrate-dependent, complicating consistent comparison of its behavior and resource cost. Aries\xspace addresses these challenges with separate task and execution specifications, {trajectory-level correlation of cross-layer telemetry}, and {a common interface} for stateful tool execution and instrumentation.

Fig.~\ref{fig:benchsuite_arch} presents the architecture of Aries\xspace. Task and execution loaders validate the specifications; the runner coordinates the harness and accesses LLM backends and sandbox substrates through adapters; and the telemetry collector merges framework events with cross-layer telemetry into a unified record. {The collector can also capture multimodal interactions, e.g., voice-to-text, across various model API and hosted backends, e.g., vLLM and SGlang, and reconstruct unified agent trajectories across these components.} This design separates task {specifications} from execution configuration, preserves cross-component provenance, and standardizes stateful tool execution across sandbox substrates.

{\textbf{Modular and configurable experimentation.}}
{Aries\xspace configures tasks and execution settings independently through separate specifications: the task specification defines instructions, benchmark-provided initial state, and success criterion; the execution specification selects the harness, model backend, sandbox, and telemetry stack.} Dedicated loaders validate the two specifications, while LLM backend and sandbox adapters bind them to a concrete execution stack without task-specific changes. {The task loader initializes these conditions, and the benchmark adapter independently evaluates the resulting task state. For example, a context-budget sweep varies only the budget, keeping the task specification, evaluator, and other execution settings fixed. Runs follow their model outputs and tool results, so actions and outcomes may differ.} This design enables controlled comparisons across configurations.

\textbf{Reconstructing cross-component trajectories.} Agent runs may contain overlapping, failed, or retried operations across independently managed components.
Aries\xspace propagates trajectory and event identifiers across model and tool boundaries and records causal and ordering metadata under a common event schema. {During reconstruction, events are grouped by run and task identity, ordered using event dependencies and component sequence metadata, and aligned with timestamped resource observations.} The telemetry collector then associates component metrics with these events, revealing where delays and resource pressure arise along the trajectory. {To reduce measurement interference, tracing is limited to correlation metadata at existing execution boundaries.}

\begin{figure}[t]
    \centering

    \vspace{-1.2em}
    \includegraphics[width=0.9\linewidth]{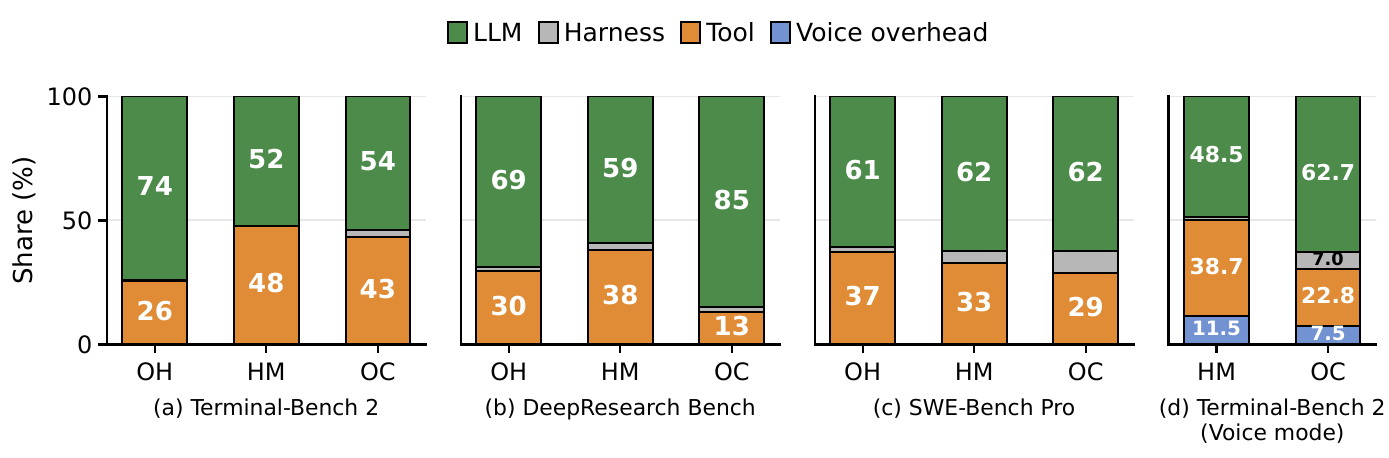}
    \vspace{-1.5em}
    \caption{End-to-end average latency breakdown {for \textsc{OpenHands} (OH), \textsc{Hermes} (HM), and \textsc{OpenClaw} (OC)}. Harness and tool execution contribute latency on par with LLM calls.
    }
    \label{fig:latency-breakdown}
    \vspace{-2.1em}
\end{figure}
\textbf{Making stateful tool execution observable and portable.}
Tool calls operate on a persistent environment whose state affects subsequent trajectory steps, while different sandbox substrates expose different lifecycle and telemetry interfaces. 
Aries\xspace standardizes environment setup, tool invocation, state continuity, result collection, and telemetry hooks, while sandbox adapters map these operations onto concrete substrates. This makes tool execution a first-class, consistently observable stage of the trajectory while supporting different sandbox technologies.

{Together, these abstractions allow Aries\xspace to re-instantiate the same controlled experimental conditions across runs, while recording variation introduced by external dependencies in the trajectory.}

\vspace{-.3em}
\section{Why Is Modern Cloud AI Infrastructure Insufficient for Agents?}

\label{sec:vision-statements}

Using Aries\xspace, we evaluate whether request-centric modern cloud AI infrastructure
is sufficient for agent serving. Specifically, we ask which components determine the end-to-end critical path, how long-lived trajectory context affects serving efficiency and accuracy, and how resources are utilized across LLM backends and tool sandboxes.
% , and the adequacy of existing tool-isolation mechanisms. 
We first analyze an eight-hour workday production trace, including request lengths from ten LLM-serving instances and one-minute CPU and memory samples from 100 agent sandboxes, to establish high-level workload behavior at deployment scale. As the operational traces do not expose synchronized cross-layer events,
{we then use Aries\xspace to study the corresponding behaviors under controlled configurations while preserving the benchmark task and evaluator.}

For this controlled analysis, Aries\xspace integrates \textsc{OpenHands}~\cite{wang2026openhands}, \textsc{Hermes Agent}~\cite{hermesagent2026}, and \textsc{OpenClaw}~\cite{openclaw2026}, and evaluates serving tasks from SWE-Bench Pro~\cite{deng2025swepro}, Terminal-Bench 2~\cite{tbench_2025}, and DeepResearch Bench~\cite{du2025deepresearch}. {All controlled experiments use Docker container-based tool sandboxes.} We sample 20 tasks from each benchmark to achieve 50\% task success rate (important for the accuracy studies) with OpenClaw and repeat each task five times. {To validate multimodal support, we convert Terminal-Bench 2 tasks to speech, transcribe them with OpenAI, and execute them with a locally served model.} Unless otherwise stated, experiments use \textsc{Qwen3.6-35B-A3B-FP8}~\cite{qwen36}, served locally by \textsc{SGLang}~\cite{zheng2024sglang} on a 96-core node with one NVIDIA H100 GPU with 94\,GB of HBM. 

\vspace{-1em}
\subsection{Harnesses \& Tools Matter as Much as Models}
\label{sec:metric-decoupling}

Existing LLM-serving engines optimize token-centric metrics such as TTFT and TPOT~\cite{zhong2024distserve,kwon2023pagedattention,zheng2024sglang}. These metrics omit the host-side intervals during which the harness processes model outputs or waits for tool execution, even though the agent cannot advance until these steps complete. 
\begin{figure}[t]
    \centering
\vspace{-1em}
    \includegraphics[width=0.95\linewidth]{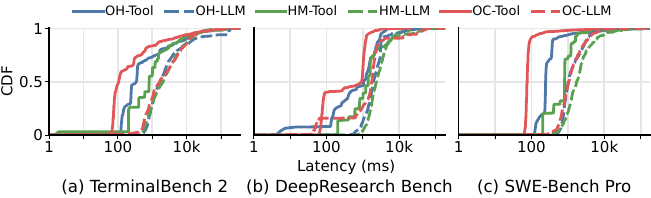}
\vspace{-1.2em}

    \caption{
        Latency distributions for LLM inference vs. tool execution. Tool calls exhibit a severe long tail spanning multiple orders of magnitude.
    }
\vspace{-2em}
    \label{fig:tool-latency-cdf}
\end{figure}
As the per-step timing information is too sensitive to collect in the commercial cluster, we base the analysis of the utility of existing metrics on the setup with public datasets and harnesses 
which record these events under a shared trajectory identifier and time base.

Our profiling highlights two key system insights:
    \textbf{(1) Critical-Path Shift to Harness and Tool Execution:} 
    As shown in Fig.~\ref{fig:latency-breakdown}, tool execution is an important latency contributor,     contributing from 13\% to 48\% of the total latency across datasets and harnesses, while reaching 48\% for Hermes Agent on Terminal-Bench 2. 
    {In addition}, the harness spends up to 9\% of end-to-end time on orchestration and other host-side processing. {For voice-mode Terminal-Bench 2 tasks, voice processing further contributes 7.5\%$\sim$11.5\%. These host-side delays increase the interval between consecutive LLM calls and may prolong the residency of trajectory state in GPU memory.}
    \textbf{(2) Long-Tailed Latency Profiles:} Fig.~\ref{fig:tool-latency-cdf} 
    shows that while LLM latency remains important, tool calls exhibit a severe long tail that spans multiple orders of magnitude. Together, these results show that these massive host-side delays frequently dominate execution time, meaning that solely optimizing accelerator token generation cannot resolve the actual agent bottlenecks.

\noindent \textbf{Takeaway 1:} 
\textit{End-to-end task duration depends heavily on harness and tool sandbox execution, rather than model inference alone. 
}

\begin{figure}[t]
\captionsetup[subfloat]{captionskip=1pt}
\centering
\vspace{-1.5em}
\subfloat[Context distribution]{
    \includegraphics[width=0.44\linewidth]{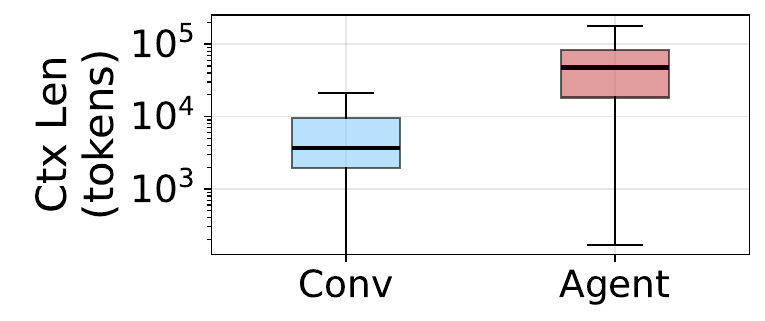}
    \label{fig:ant_context_capacity}
  }
  \hfill
  \subfloat[Resident capacity]{
    \includegraphics[width=0.44\linewidth]{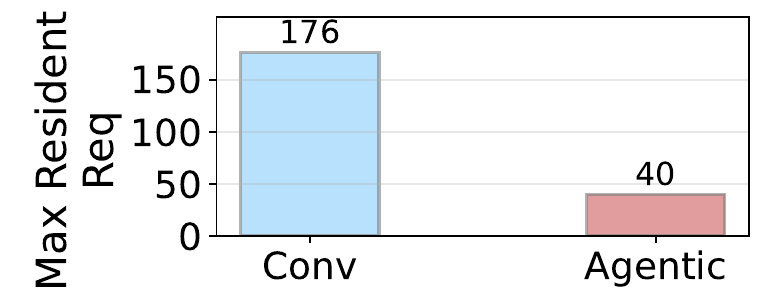}
    \label{fig:ant_resident_sessions}
  }
  \vspace{-1em}
\caption{
Context length and resident capacity comparison across conversation and agent requests in the production cluster. Agent workloads consume larger per-request KV footprints than conventional serving, sharply reducing the number of processed requests in parallel by an LLM backend.
}
\label{fig:long_context_capacity}
  \vspace{-.5em}
\end{figure}

\subsection{Long Context Dictates Efficiency \& Accuracy}
\label{sec:eff_acc_tradeoff}

In conventional LLM serving, context lifetimes are typically scoped to the request-response cycle, allowing engines to reclaim KV cache pages post-query unless explicitly retained or offloaded for reuse~\cite{kwon2023pagedattention,zheng2024sglang,zhong2024distserve,hu2024memserve,qin2024mooncake,liu2025lmcache}. Conversely, long-horizon agents generate monotonically growing histories that persist across the entire trajectory. While existing frameworks apply semantic compression to fit system limits~\cite{rasmussen2025zep,xu2026amem,chhikara2025mem0}, unmanaged context expansion still locks up physical GPU token pools and reduces serving concurrency~\cite{li2025continuum,kariyappa2026sidequest}. Retaining excessive history can increase tail delay, while aggressive compression may discard information required by later steps and reduce task accuracy~\cite{kang2025acon,kariyappa2026sidequest}. Resolving this accuracy-efficiency tension requires shifting from passive memory allocation to trajectory-aware state management that balances the resource cost and future utility of retained context.

We first compare the context lengths observed in agentic serving and conventional inference, both retrieved from the production platform. Then, we characterize the trade-offs associated with the context budget and how they affect the accuracy and efficiency of agentic serving with the public harnesses and datasets. In the latter analysis, across all evaluated benchmarks, we sweep the context budget for both \textsc{OpenClaw} and \textsc{Hermes Agent} to determine when additional retained history stops providing meaningful accuracy gains. We then quantify the cost of context retention when serving 32 concurrent active agentic sessions with a 256K context size, sampling backend resource utilization once per second. {\textsc{OpenHands} exhibits the same qualitative trends as \textsc{Hermes} as the context window size varies.}

\begin{figure}[t]
    \centering
    \vspace{-1em}
    \includegraphics[width=1\linewidth]{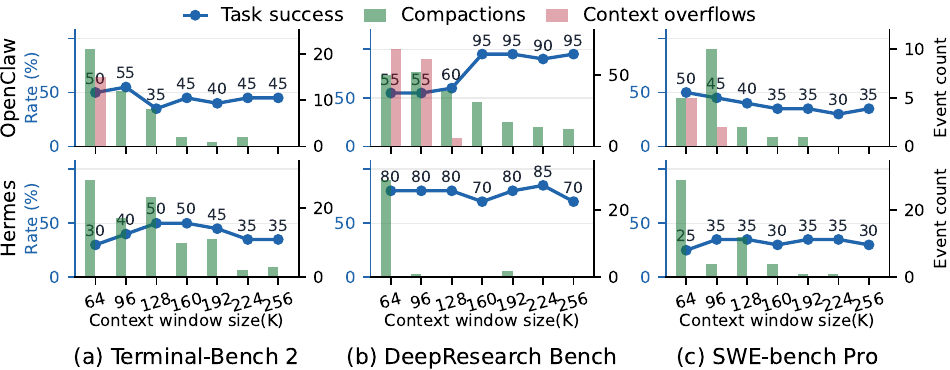}
    \vspace{-1.5em}
    \caption{Compaction and context overflow counts with different context window sizes, and the corresponding task success rate. Larger windows ease overflow/compaction pressure, but task success plateaus past workload-specific thresholds.}
    \vspace{-1em}
\label{fig:context_success}
\end{figure}

\begin{figure}[t]
    \centering
    \vspace{-1.2em}
    \subfloat[Concurrency of active trajectories (Act.), resident LLM-engine sessions (Res.), and queued sessions (Queue).\label{fig:concurrency}]{
      \includegraphics[width=0.3\linewidth]{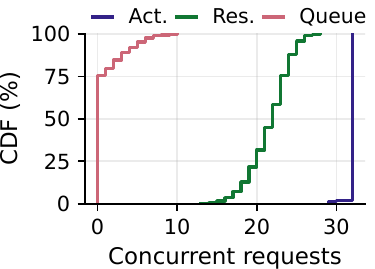}
    }
    \hfill
    \subfloat[Resource pressure from KVC pool usage(KV), SM activity(SM) and occupancy(Occ), and HBM activity(HBM).\label{fig:resource_pressure}]{
      \includegraphics[width=0.3\linewidth]{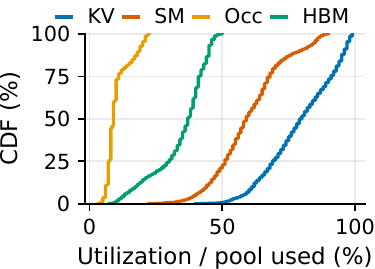}
    }
    \hfill
    \subfloat[Request waiting time measured by TTFT and SGLang queue time (Queue).\label{fig:waiting_time}]{
      \includegraphics[width=0.3\linewidth]{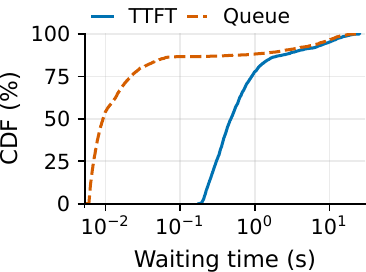}
    }
    \vspace{-1em}
    \caption{ CDFs over the one-hour window of the agent serving. The workload sustains high concurrency and KV pressure, while long waits are concentrated in the request tail.}
    \vspace{-.5em}
    \label{fig:concurrency_resource_waiting}
\end{figure}

We make three key observations: 
\textbf{(1) Trajectory State Amplification:} Production traces reveal substantially larger agent contexts than those observed in conventional inference workloads, which matches our results with the public datasets and harnesses. As quantified in Fig.~\ref{fig:long_context_capacity}, the average agent context consumes a substantially larger KV footprint than conventional inference. Under a fixed 25GB budget with the Qwen3.6 model, this state amplification reduces the maximum resident capacity from over 176 concurrent serving requests to merely 40---a $4.4\times$ reduction.
\textbf{(2) Context Sufficiency Thresholds:} Using the open datasets and harnesses, we show that long-context retention improves accuracy only up to a task-specific threshold, beyond which additional history, albeit consuming more memory, provides little to no accuracy benefits. As shown in Fig.~\ref{fig:context_success}, \textsc{OpenClaw} on DeepResearch Bench reaches peak success (95\%) at 160K context, coinciding with the elimination of context overflows. Conversely, Terminal-Bench 2 and SWE-Bench Pro plateau or degrade once overflow pressure is removed, performing best at small or mid-size windows rather than the maximum context size.
\textbf{(3) Capacity-induced Batch Limitation:} 
our results with public datasets further show that large retained state severely constrains serving capacity. 
Fig.~\ref{fig:concurrency} and ~\ref{fig:resource_pressure} show that memory pressure limits effective batching despite a 32-session backlog; token-pool utilization peaks at 97\%, while the engine sustains a median of only 22 concurrent requests. Under this pressure, the scheduler preempts active sessions, and the $p_{95}$ internal queueing delay reaches 7.9s (Fig.~\ref{fig:waiting_time}).

\noindent 
\textbf{Takeaway 2:}
\textit{
Retaining context beyond its task-dependent utility yields diminishing accuracy benefits while increasing KV-cache occupancy, limiting batching, and prolonging queueing delays. The system must therefore actively manage trajectory state to balance task accuracy against serving efficiency.
}

\subsection{Tool Sandboxes Must Be Elastic}
\label{sec:elastic-cpu-mem}
\vspace{-.5em}

Agent serving introduces CPU-based processing to the critical path: it alternates between idle intervals during LLM inference and short CPU/memory bursts during tool execution. Static peak resource provisioning avoids throttling these tool calls but leaves resources stranded a significant portion of the time. Cloud systems such as AWS Lambda MicroVMs~\cite{aws2026lambdamicrovms} and Google Agent Sandbox~\cite{google2026gkeagentsandbox} attempt to reduce waste by applying a \textit{stateful scale-to-zero} approach via sandbox snapshot capture and restore (C/R). However, agentic tool execution patterns lead to a high number of capture-restore cycles, incurring additional costs, as we show below.

\begin{figure}[t]
  \centering\captionsetup[subfloat]{captionskip=-.2em}
  \vspace{-2em}
  \subfloat[Harness CPU]{
    
    \includegraphics[width=0.40\linewidth]{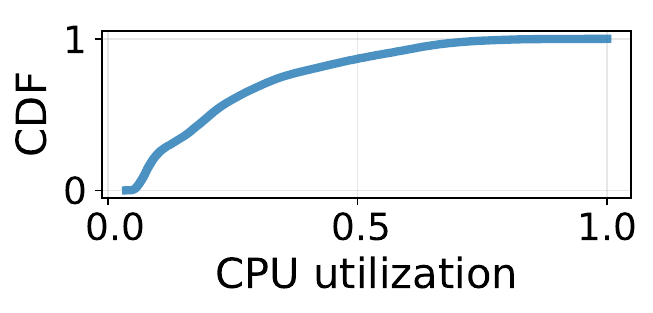}
    \label{fig:ant_cpu_utilization_cdf}
  }
  \hfill
  \subfloat[Harness memory]{
    \includegraphics[width=0.40\linewidth]{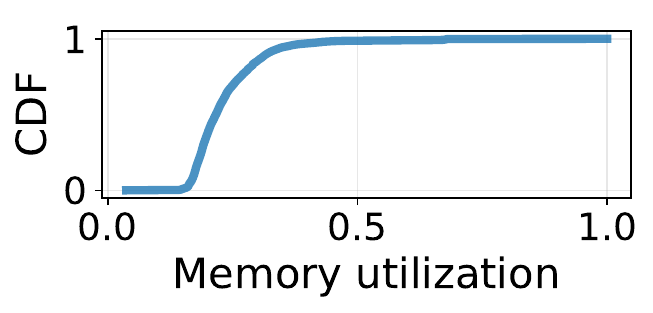}
    \label{fig:ant_memory_utilization_cdf}
  }
  \vspace{-1em}
  \caption{ CDFs of Harness CPU and memory utilization 
  aggregating measurements taken every minute from the production cluster. The values are normalized for trace anonymization purposes.
  \vspace{-.5em}
  }
  \label{fig:ant_utilization_cdf}
\end{figure}

{The harness manages the agent loop and issues tool requests, while the tool sandbox maintains the task environment and executes the requested commands.}
We collect CPU and memory utilization from over 100 sandboxes on the same commercial platform, each co-hosting an OpenClaw-style harness and its tool calls.
To confirm the trend and for in-depth analysis, we also collect CPU and memory utilization of harness and tool sandboxes in a controlled environment every second and analyze their distributions for \textsc{OpenClaw} and \textsc{Hermes} using the cloud DeepSeek V4 Flash API, which is production-grade and highly stable for tool call generation accuracy.
We then analytically replay the recorded tool-call traces against the AWS Lambda MicroVMs pricing model~\cite{aws2026lambdapricing} under different sandbox keepalive policies as in serverless deployments~\cite{shahrad2025serverless, fuerst2021faascache}.
We report the instance cost (CPU and memory usage), the snapshot C/R cost, and their sum. The \(\infty\) keepalive point corresponds to the persistent sandbox. DeepResearch is excluded from this comparison due to the tools' low CPU usage. 

\begin{figure}[t]
\captionsetup[subfloat]{captionskip=-.3em}
  \vspace{-1.9em}
  \centering
  \subfloat[CPU]
  {
    \includegraphics[width=\linewidth]{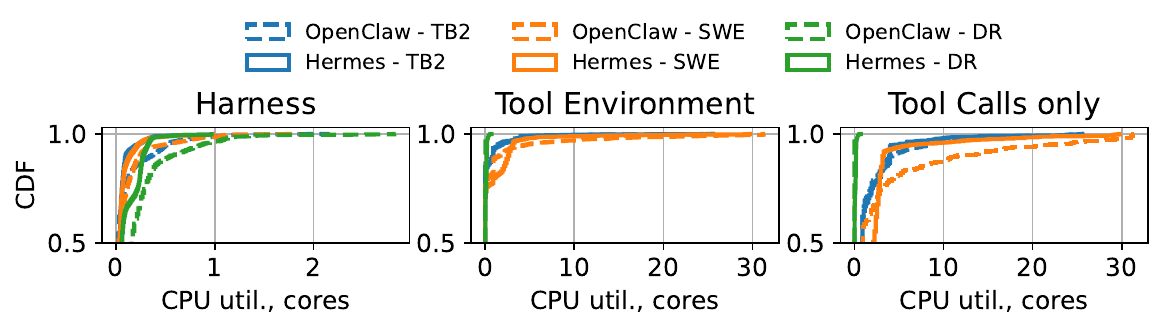}
    \label{fig:cpu_utilization_cdf}
  }
  \vspace{-1.2em}

  \subfloat[Memory]
  {
    \includegraphics[width=\linewidth]{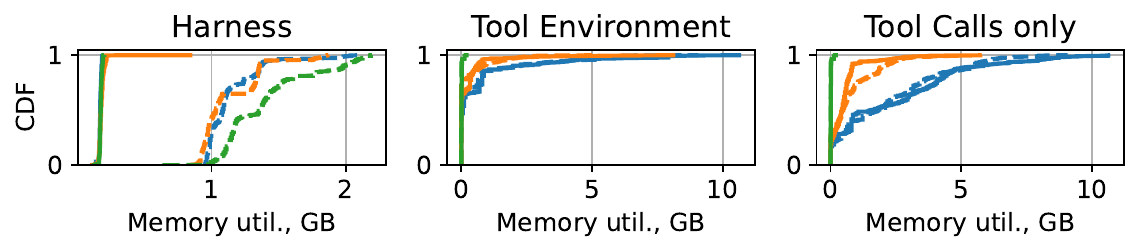}
    \label{fig:memory_utilization_cdf}
  }
  \vspace{-1.2em}

  \caption{Tool and harness CPU and memory utilization CDFs aggregating measurements taken every second.
  The tool environment is mostly idle, with high CPU utilization spikes during tool calls.
  }
  \vspace{-.5em}
  \label{fig:utilization_cdf}
\end{figure}

\begin{figure}[t]
  \centering
  \vspace{-.5em}
  
  \includegraphics[width=1.\linewidth]{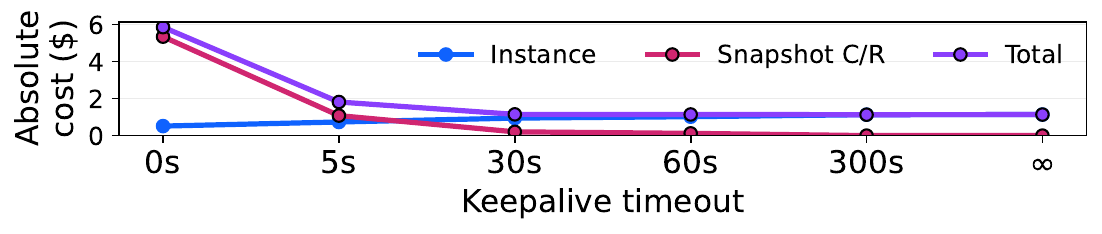}
  \vspace{-3em}
  \caption{Absolute customer-visible cost of snapshot-based stateful scale-to-zero under different keepalive timeouts based on AWS Lambda MicroVM billing model.
  Short keepalive reduces instance cost but triggers frequent snapshot C/R, while longer keepalive reduces C/R by keeping the sandbox resident for more idle time.
  \vspace{-1.5em}
  }
  \label{fig:cost_breakdown}
\end{figure}

Fig.~\ref{fig:ant_utilization_cdf} shows that CPU compute and memory utilization in the production setting is spiky. 
We reproduce this with the public datasets and harnesses (Fig.~\ref{fig:utilization_cdf}),
showing that agent tool sandboxes spend most of their trajectories at low utilization, yet exhibit sharp spikes in CPU and memory demand during tool execution. Across harnesses and workloads, CPU and memory remain idle for more than 80\% and 50\% of the time, respectively. At the same time, the tool-call-conditioned distributions show much more intensive resource use: the portion of time with idle resources drops below 20\% for both CPU and memory.
Fig.~\ref{fig:cost_breakdown} shows that both users and cloud providers have an incentive to reduce keepalive to lower their costs: users avoid paying for idle sandboxes, while providers reclaim stranded capacity sooner. Yet current snapshot-based mechanisms and pricing prevent this alignment: if tool sandboxes are torn down after each tool call, their cost falls by 65\%, but repeated snapshot C/R raises the total customer cost to 4.9$\times$ compared to the persistent baseline. Keeping tool sandboxes for longer reduces C/R overhead but increases idle time. Thus, today's system forces a false choice between paying for idle capacity and paying for repeated snapshots and restores.
However, these results suggest that the total cost can be reduced by 3$\times$ in an ideal system compared to the persistent-sandbox baseline.

\noindent \textbf{Takeaway 3:}
\textit{
Agent sandboxes are mostly idle but bursty when tools execute. Both users and providers should benefit from idle resource reclamation, yet current state-management mechanisms and pricing make aggressive suspension uneconomical. 
}

\vspace{-.4em}

\vspace{-.2em}
\section{Future Research Directions}

\begin{figure}[t]
    \centering
    \vspace{-.6em}
    \includegraphics[width=1.\linewidth]{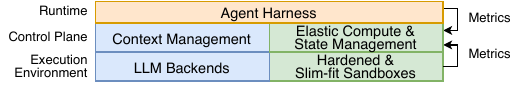}
    \vspace{-2.5em}
    \caption{Agent-native AI serving system vision.
    }
    \vspace{-1.5em}
    \label{fig:future-dir}
\end{figure}

Our results suggest that agent-serving systems should coordinate optimization decisions across the trajectory rather than manage model inference, context state, and tool execution independently. As illustrated in Fig.~\ref{fig:future-dir}, this requires jointly optimizing task progress, capacity, and resources through three co-designed directions across the Runtime, Control, and Execution layers: (1) trajectory-capturing metrics spanning all layers; (2) control plane for trajectory-aware context management; and (3) elastic tool execution management via sandbox control.
% ; and (4) minimizing the tool sandbox attack surface.

\vspace{-.5em}
\subsection{Trajectory-capturing Metrics}
\begin{figure}[t]
    \centering
    \vspace{-1.2em}
    \includegraphics[width=\linewidth]{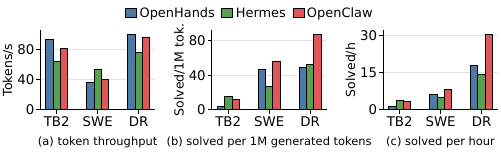}
    \vspace{-2.5em}
    \caption{Token throughput, solved tasks per 1 million generated tokens, and solved tasks per hour. Token throughput can diverge from task goodput and execution efficiency in agent serving.
    }
    \vspace{-2em}
    \label{fig:task-goodput}
\end{figure}

Future agent-serving metrics should continuously quantify how system execution contributes to task progress. As illustrated in Fig.~\ref{fig:task-goodput}, evaluation should connect end-to-end outcomes, such as task success and completion rate, with low-level measures of execution speed and cost, including latency, generated tokens, and resource consumption. Task goodput and solved tasks per generated token provide useful endpoint metrics, but they become available only after a trajectory completes and cannot guide runtime decisions.

A key research direction is therefore to derive online utility signals from trajectory events. Such signals should estimate the incremental progress contributed by each model call, harness decision, and tool interaction while accounting for its time and resource cost. They should distinguish productive actions from repeated or non-advancing steps and aggregate into end-to-end measures of task accuracy, execution speed, and system cost. This connection would allow low-level system optimizations to be evaluated by their actual contribution to agent task completion.

\vspace{-.8em}
\subsection{Control Plane for Trajectory-Aware Context Management}
Our measurements in \S\ref{sec:eff_acc_tradeoff} reveal a workload-dependent trade-off between context sufficiency and serving capacity.
Additional context increases KV-cache pressure and queueing delay, even when it provides only a small improvement in task success.
This motivates a trajectory-aware control plane that monitors task and memory states, determines how context should be retained and placed, and directs the LLM backend to perform context compaction, KV-cache compression, or state transfer.
\textbf{(1) Adaptive Context Retention.} The control plane should estimate the future utility of context as a trajectory evolves and request compaction or compression when the expected accuracy benefit of retained state no longer justifies its memory cost. Such decisions must remain conservative enough to preserve information needed by later model invocations.
\textbf{(2) Hierarchical State Placement.} The control plane should coordinate long-lived context placement across accelerator and lower-cost memory tiers rather than leaving all state resident on the accelerator. It should trigger KV-cache migration when the released accelerator capacity outweighs the transfer and restoration overhead.

\vspace{-.5em}
\subsection{Control Plane for Agent-Aware Elasticity}

Our characterization in \S\ref{sec:elastic-cpu-mem} shows that host-side resource demand is highly intermittent.
Static provisioning wastes capacity, while conventional, horizontal autoscaling mechanisms are too expensive at such granularity.
Future agent-serving platforms therefore require an elasticity control plane designed around model-driven, stateful execution.
\textbf{(1) Vertical, Model-Aware Compute Management.}
Research should explore how signals from the model, harness, and runtime can guide timely vertical scaling of individual sandboxes to satisfy the identified volatile demand for memory and core count fluctuations. Key questions include which signals reliably predict demand, how uncertainty should be handled, and how resources should be allocated across competing trajectories. 
\textbf{(2) State Management for Elastic Execution.}
Because sandboxes retain files, processes, caches, and network state, elasticity requires decoupling execution state from physical placement and resource allocation. Research is needed on what state must remain resident, what can be externalized or reconstructed, and how to preserve consistency while resizing, suspending, or relocating sandboxes.

\vspace{-.5em}
% \subsection{Sandboxes with a Minimal Attack Surface}
% \vspace{-0em}

% AI agents are increasingly capable of autonomously finding and exploiting vulnerabilities.
% Thus, cloud infrastructure should anticipate increasingly sophisticated adversarial workloads.
% \textbf{(1) Defense-in-Depth and Minimizing Attack Surface.}
% {Sandboxes should be tailored to the workload to reduce the attack surface by restricting access to tools, networks, and filesystems, and leverage the host-side VM-level isolation with guest-side kernel-level sandboxing.}
% For instance, workloads such as deep research do not need direct \texttt{bash} access, but can be granted a sandboxed JavaScript runtime within the guest.
% \textbf{(2) Formal Security Guarantees.}
% VM isolation moves the trust boundary to the VMM and its host-facing interfaces. In the long term, migrating to formally verified systems, such as seL4~\cite{klein2009sel4}, as a host hypervisor or guest kernel would improve security guarantees at the cost of a major system overhaul.

% \input{section/06_related}
\section{Conclusion}

Agent serving turns conventional LLM serving into long-lived, stateful trajectories spanning model execution, harness logic, and sandboxes. We present Aries\xspace, a telemetry framework that combines reproducible experiments with production traces to characterize task and system behavior. 
Our findings motivate trajectory-capturing metrics and control planes for context management and agent-aware elasticity.
% , and sandboxes with minimal attack surfaces.

\vspace{-.5em}
\begin{acks}
We thank the anonymous reviewers for invaluable feedback and shepherding.
We also thank Charly Castes for valuable security discussions.
This research is supported by the Ministry of Education, Singapore, under its Academic Research Fund Tier 2 (MOE-T2EP20124-0002) and Ant Group.
\end{acks}

\bibliographystyle{ACM-Reference-Format}
\bibliography{agent}

\end{document}